\documentclass[11pt,njp]{iopart}

\newcommand{\braket}[1]{\ensuremath{\left\langle{#1}\right\rangle}}
\newcommand{\ket}[1]{\left\vert#1\right\rangle}

\newcommand{\bra}[1]{\left\langle#1\right\vert}

\usepackage{graphicx,subfigure}
\usepackage{times}

\begin{document}

\title{\sf \bfseries Dynamics of atom-atom correlations in the Fermi problem}

\author{\sf \bfseries Massimo Borrelli$^{1}$, Carlos Sab{\'i}n$^{2}$, Gerardo Adesso$^{3}$, Francesco Plastina$^{4,5}$, Sabrina Maniscalco$^{1,6}$}

\address{$^{1}$SUPA, EPS/Physics, Heriot-Watt University, Edinburgh, EH14 4AS, United Kingdom\\
$^{2}$Instituto de F\'isica Fundamental, CSIC, Serrano 113-B, 28006 Madrid, Spain\\
$^{3}$School of Mathematical Sciences, The University of Nottingham, University Park, Nottingham NG7 2RD, United Kingdom\\
$^{4}$Dipartimento di Fisica, Universit\`a della Calabria, 87036 Arcavacata di Rende (CS), Italy\\
$^{5}$INFN-Gruppo Collegato di Cosenza, Italy\\
$^{6}$Turku Center for Quantum Physics, Department of Physics and Astronomy,
University of Turku, FIN20014, Turku, Finland}
\ead{mb325@hw.ac.uk}
\begin{abstract}
We present a detailed perturbative study of the dynamics of
several types of atom-atom correlations in the famous Fermi problem. This is
an archetypal model to study micro-causality in the quantum domain
where two atoms, the first initially excited and
the second prepared in its ground state, interact with the
vacuum electromagnetic field. The excitation can be transferred
to the second atom via a flying photon and various kinds of quantum
correlations between the two are generated during this process.
Among these, prominent examples are given by entanglement,
quantum discord and nonlocal correlations. It is the aim of this paper to analyze the role
of the light cone in the emergence of such correlations.

\end{abstract}


\section{Introduction}
Since its original conception, the two-atom Fermi problem
\cite{fermi} has been the subject of an intense academic debate.
Stated in very simple terms, this {\itshape gedanken} experiment
goes as follows: suppose there are two two-level atoms (qubits),
say $A$ and $B$, spatially separated by a distance $r$ and
interacting independently with a the (multi-mode) quantized
electromagnetic field, initially in the vacuum state. The  atom
$A$ is in an excited state $|e\rangle$ whereas the atom $B$  is in
its ground state $|g\rangle$. At some time $t_{0}$ the atom $A$
emits a photon. The original question by Fermi was then the
following: as long as the two atoms are causally disconnected, is
the excitation probability of $B$ independent on the presence of
$A$? This question points at the very foundations of quantum
mechanics: do quantum mechanical probabilities respect
micro-causality? Over the past years a considerable amount of
literature has dealt with this problem and several perturbative
solutions have been proposed
\cite{biswas,heger,buch,power,milonni}, sometimes even presenting
opposite conclusions. Recently, in Ref.\cite{carlos1}, a
non-perturbative proof of strict causality in the Fermi problem
has been finally given along with an explanation of why the
existence of correlations outside the light-cone connecting the
two atoms is not in contrast with micro-causality. In fact, in a
previous paper \cite{carlos2}, some of the same authors had
already studied the dynamics of concurrence \cite{concurrence} and
found that it starts increasing just before $(t-t_{0})=r/c$ by a
very tiny amount.

In this paper, motivated by such results, we take a step further
and investigate the dynamics of other types of atom-atom
correlations. In particular, besides extending the analysis of
entanglement dynamics, we study the time evolution of geometric
quantum discord and (classical) connected correlation functions.
Quantum discord was first introduced by Ollivier and Zurek
\cite{zurek,anvedi} as a novel measure of the quantumness of
correlations. The idea behind it is conceptually very simple.
Quantum discord is indeed defined as the discrepancy, in the
quantum regime, between two classically equivalent definitions of
mutual information. It is believed to capture a more general type of
quantum correlation than entanglement, in the sense that quantum
states with zero entanglement but non-zero quantum discord do exist
(see e.g.~\cite{ferraro,piani,streltsov,plastica,gerardo,pianinew}).
Unfortunately, the computation of  quantum discord implies solving
a rather complicated minimization problem and, although
considerable improvements have been made over the last years
\cite{reviewmodi}, analytical results are available for very few
cases only \cite{luo,davide,discordgaussiano}. In order to
overcome such computational issues, alternative indicators of
general quantum correlations have been recently introduced, mostly
based on distance-based approaches \cite{vedral,kavan}. On the
other hand, connected correlation functions provide a statistical
quantifier of the (classical) correlations extractable during a
joint measurement of the two atoms
\cite{localizable1,localizable2,arealaw}.

The results reported in this article have a double merit: on one
hand they give a more complete picture of the dynamics of
correlations in the two-atom Fermi problem. In fact, to the best
of our knowledge, this is the first study of quantum discord and
more general correlation dynamics in such a physical model, where
an exact solution of the dynamics is still missing. As it turns out, all the
types of correlations we consider have a nice physical
interpretation in terms of a few relevant physical processes of
the dynamics. On the other hand our results suggest a way to
detect atom-atom correlations outside the light cone, that is,
when the two atoms are causally disconnected. It is important to
remark that our calculations are performed in the framework of
time-dependent perturbation theory and they are exact and
consistent up to the second order in the coupling constant.
However, this is certainly not a problem since, as stated above,
strict causality in the model has been analytically proven with no
usage of perturbation theory \cite{carlos1}. Moreover, the time
interval under scrutiny falls within the limits of validity of our
approach provided the two atoms are not that distant.

\section{The Model}
\label{model}
We will consider a one-dimensional physical setup. A pair of
two-level superconducting qubits (artificial atoms) $A$ and $B$,
separated by a fixed distance $r$, interact with an
electromagnetic field propagating along the open transmission line
connecting them. We name the atomic levels as
$\{|g\rangle,|e\rangle\}$ and we assume the following multimode
structure for the field
\begin{equation}
V(x)=\int dk \sqrt{N\omega_{k}}\left[e^{ikx}a_{k}+e^{-ikx}a^{\dagger}_{k}\right]
\label{field}
\end{equation}
where $N$ is a normalization factor which may accommodate
different circuit QED architectures, the dispersion relation is
linear $\omega_{k}=\upsilon |k|$, and $a_{k}, a^{\dagger}_{k}$ are
the usual annihilation and creation operator satisfying boson
commutation relations $[a_{k},a^{\dagger}_{k'}]=\delta_{k,k'}$. We
define $\Omega_J=\omega_{Je}-\omega_{Jg}$ ($J=A,B$) the energy
separations between the qubit levels  and we assume the qubits to
be much smaller than the relevant wavelengths $\lambda_J=
\upsilon/(\Omega_J/(2\pi))$, $\upsilon$ being  the propagation
velocity of the field quanta which in this scheme depends on the
microscopic details of the model. Specifically,
$\upsilon=1/\sqrt{cl}$, $c$ and $l$ being the capacitance and
inductance per unit length respectively. A typical value is
$\upsilon=1.2\cdot10^8 m/s$ \cite{casimirwilson}. Under these
conditions the Hamiltonian, $H = H_0 + H_I,$  splits into a free
part for the qubits and the field
\begin{equation}
 H_0 = \frac{1}{2}\hbar(\Omega_A\sigma^z_A + \Omega_A\sigma^z_A) + \int_{-\infty}^{\infty}dk\, \hbar\omega_k  a^{\dagger}_ka_k \label{h0}
 \end{equation}
and a point-like interaction between them
\begin{equation}
  H_I = - \sum_{J=A,B} d_J\,V(x_J)\,\sigma_J^x\label{c}
\end{equation}
Here $x_J$  are the fixed positions of the atoms, and $d_J\,
\sigma^x_J$ comes from a dimensional reduction of the matter-
radiation interaction hamiltonian with two-level atoms and the
electromagnetic field. We will consider the following  initial
state
\begin{equation}
|\psi(0)\rangle = |eg0\rangle,\label{eq:initialstate}
\end{equation}
where only qubit $A$ is excited, while $B$ and the field remain in
their ground and vacuum states, respectively. We use the formalism
of perturbation theory  up to the second order and beyond Rotating
Wave Approximation \cite{carlos2}  and trace over the field
degrees of freedom to obtain the corresponding two-qubit reduced
density matrix $\rho_{\rm X}$ evaluated at $t$.
In the interaction picture with respect to the free Hamiltonian
$H_0,$ the system evolves during a lapse of time $t$ into the
state
\begin{equation}
  \ket{\psi(t)} = {\cal T}[e^{-i \int_0^tdt' H_I(t')/\hbar}]\ket{eg}\otimes\ket{0},\label{c}
\end{equation}
${\cal T}$ being the time ordering operator. Up to second order in perturbation theory the final state can be written as
\begin{eqnarray}
  \ket{\psi(t)}  &= &  \left[(1+A)\ket{eg} + X\ket{ge}\right]\otimes\ket{0} +
  \label{wavefunction} \nonumber \\
 &&  (U_A\ket{gg} + V_B\ket{ee})\otimes\ket{1} \nonumber \\ && + (F\ket{eg} +  G \ket{ge})\otimes\ket{2} + {\cal O}(d^3).
\end{eqnarray}
The coefficients for the vacuum, single-photon, and two-photon states,
are computed using the action $(\alpha=A,B)$
\begin{eqnarray}
  \mathcal{S}^+_\alpha \! = \! - \frac{i}{\hbar}
  \int_0^t
  e^{i\Omega t'}\braket{e_\alpha|d\sigma^x_\alpha|g_\alpha} V(x_\alpha,t') dt'
  = -(\mathcal{S}^{-}_\alpha)^\dagger\label{f}
\end{eqnarray}
among different photon number states $\ket{n}, n=0,1,2\ldots$,
being $\ket{n}\bra{n}=\frac{1}{n!}\int dk_1....\int
dk_n\ket{k_1...k_n}\bra{k_1...k_n}$ and
$\ket{k}=a_k^{\dagger}\ket{0}$. Among the various terms present
here, the only one containing an effective coupling between $A$
and $B$ is
\begin{equation}
  X = \langle0|T(\mathcal{S}^+_B \mathcal{S}^-_A)|0\rangle . \label{exchange}
\end{equation}
This includes photon exchange only inside the light cone, $vt>r,$
and vacuum fluctuations for all values of $t$ and $r$, being
$r=x_B-x_A$ the distance between the qubits.  The remaining terms
are
\begin{eqnarray}
A & \!\! = \!\! & \frac{1}{2}\bra{0}T(\mathcal{S}_A^+ \mathcal{S}_A^- +
\mathcal{S}_B^-\mathcal{S}_B^+)\ket{0}\label{e}\\
U_A & \!\! = \!\! & \bra{1}\mathcal{S}^-_A\ket{0},
V_B = \bra{1}\mathcal{S}^+_B\ket{0} , \nonumber\\
F & \!\! = \!\! & \frac{1}{2}\bra{2}T(\mathcal{S}_A^+ \mathcal{S}_A^-
+\mathcal{S}_B^-\mathcal{S}_B^+)\ket{0} \! , \, G = \bra{2}T(\mathcal{S}^+_B \mathcal{S}^-_A)\ket{0}.\nonumber
\end{eqnarray}
Here, $A$ describes intra-qubit radiative corrections, while $U_A,
V_B, F$ and $G$ correspond to single-photon emission events by one
or more qubits.
The coefficients in Eq.~(\ref{wavefunction}) will be computed
analytically as a function of two dimensionless parameters, $\xi$
and $K$. The first one, $\xi=\upsilon t/r$, is a dimensionless
time variable; the time $\xi=1$ corresponds to the light-cone,
which separates two different spacetime regions, before and after
photons can be exchanged. The second parameter is a dimensionless
coupling strength
\begin{equation}
K=\frac{4d^2N}{\hbar^2 \upsilon} =
2\left(\frac{g}{\Omega}\right)^2\label{g}.
\end{equation}
Note that the qubit-line coupling $g=d\sqrt{N\Omega}/\hbar$
corresponds to the qubit-cavity coupling that appears by taking
the same transmission line and cutting it in order to have a
length $L=\lambda$ (thus creating a resonator). This formulation
has the advantage of being valid both for inductive and capacitive
coupling, the details being hidden in the actual expressions for
$d$ and $N$.
Tracing over the states of the field, we arrive at the following
reduced density matrix
\begin{eqnarray}
\rho_{\rm X}=\frac{1}{c}\left( \begin{array}{c c c c}
\rho_{11}&0&0&\rho_{14} \\
0&\rho_{22}&\rho_{23}&0\\
0&\rho_{23}^*&\rho_{33}&0\\
\rho_{14}^*&0&0&\rho_{44}
\end{array}\right) , \label{state}
\end{eqnarray}
representing the two-qubit state in the basis formed by $\ket{ee},$
$\ket{eg},$ $\ket{ge},$ and $\ket{gg}.$ The coefficients with the
leading order of neglected contributions are
\begin{eqnarray}
\rho_{11}&=&|V|_B^2+\mathcal{O}(d^4),~
\rho_{22}=1+2\mathrm{Re}(A)+\mathcal{O}(d^4)\nonumber\\
\rho_{33}&=&|X|^2+|G|^2+\mathcal{O}(d^6),~
\rho_{44}=|U|_A^2+\mathcal{O}(d^4) \nonumber\\
\rho_{14}&=&U_A^*V_B+\mathcal{O}(d^4)=\langle0|\mathcal{S}_A^+ \mathcal{S}_B^+|0\rangle +\mathcal{O}(d^4)\label{j}\\
\rho_{23} &=& X^*+\mathcal{O}(d^4) , \nonumber
\end{eqnarray}
and the state is normalized, $c=\sum_i \rho_{ii}.$

\section{Dynamics of Correlations}
\label{correlations}
In this section we report our results
regarding the dynamics of correlations between the two
(artificial) atoms. We investigate the time evolution of the
square root of geometric quantum discord $\sqrt{D}$ \cite{vedral},
of the entanglement as measured by the negativity $N$ \cite{neg},
and of the maximum connected correlation function $C$
\cite{localizable2} in the state $\rho_{\rm X}$. We have chosen
these three specific types of correlations (whose definitions and
properties are reported below) for four reasons:
\begin{itemize}
\item[(i)] to best relate our results with the ones reported in \cite{carlos2} by using a different measure of entanglement;
\item[(ii)] to have a more complete description of the time evolution of general quantum correlations;
\item[(iii)] to compare quantum correlations with correlations having also a classical nature;
\item[(iv)] to issue a comprehensive comparative analysis among {\it compatible} measures of different types of correlations.
\end{itemize}
The meaning of the latter point can be clarified in connection
with our choice of using $\sqrt{D}$. One might in fact question
why we are comparing different powers of geometric discord and
entanglement. The reason for this is that we want to be consistent
in the order of expansion of the perturbative analysis we have
performed. A good test to check whether this is true is provided
by a hierarchy-type relationship that exists between the three
chosen quantities for arbitrary states $\rho$ of two qubits, namely
\begin{equation}\label{hierardy}
C(\rho) \geq \sqrt{D}(\rho) \ge N(\rho)\,.
\end{equation}
The rightmost inequality in (\ref{hierardy}) was proven
analytically in \cite{gerardo}, while the leftmost one has been
verified numerically in \cite{unpube}. Notice that for pure
two-qubit states both inequalities are saturated and
(\ref{hierardy}) becomes a chain of equalities.

In our analysis, we have found no violation of the hierarchy
(\ref{hierardy}) for any range of the relevant physical parameters
characterizing the states $\rho_{\rm X}$, which serves as a
validating indication that our results are consistent up to the
second order. As a general remark, we can conclusively state that
all correlations whose time evolution we have looked at, start increasing
before the time at which the two atoms become causally connected.
However, the rate of increase changes a lot from one type of
correlation to another.

Let us now introduce the measures of correlations of interest and
then discuss the main aspects of their dynamical behavior. To this
end, the Bloch representation of generic two-qubit states $\rho$
will be useful \cite{luo}. Namely,
 \begin{eqnarray}\label{blopix}
 \rho &= \frac14 \sum_{i,j=0}^3 R_{ij} \sigma_i \otimes \sigma_j \\
 &= \frac 14\left(\textrm{I}_{1}\otimes\textrm{I}_{2}+\sum_{i=1}^3 x_i\sigma_i \otimes \textrm{I}_{2} +\sum_{j=1}^3 y_j \textrm{I}_{1}\otimes \sigma_j+\sum_{i,j=1}^3 t_{ij} \sigma _i\otimes\sigma_j\right)\,, \nonumber
 \end{eqnarray}
where $R_{ij}=\textrm{Tr}[\rho(\sigma_i\otimes \sigma_j)]$,
$\sigma_0=\textrm{I}$, $\sigma _i$ ($i=1,2,3$) are the Pauli
operators; $\vec{x}=\{x_i\}$ and $\vec{y}=\{y_i\}$ represent the
three-dimensional Bloch column vectors associated to the qubits
$A$ and $B$, respectively; and  $t_{ij}$ are the elements of the
$3 \times 3$ correlation matrix $T$.

\subsection{Geometric discord}
The geometric measure of quantum discord was first introduced in
\cite{vedral} and further investigations for two-qubit systems
have been reported in \cite{luofu,gerardo,gprl}. Given a general
bipartite $d_A\otimes d_B$ quantum state $\rho$, the (normalized)
geometric discord is defined as
\begin{equation}
D(\rho)\doteq\frac{d_A}{d_A-1}
\min_{\chi\in\Omega_{0}}||\rho-\chi||_{2}^{2}
\label{gd}
\end{equation}
where $\chi$ is a so-called classical-quantum state belonging to
the set of zero-discord states $\Omega_{0}$, $\chi= \sum_i p_i
|i\rangle\langle i| \otimes \varrho_{i B}$ and
$||P-Q||_2^{2}=\textrm{Tr}(P-Q)^{2}$ is the squared
Hilbert-Schmidt distance between a pair of operators $P, Q$. We
can look at geometric discord as the minimum disturbance that
would be induced in the system after a projective measurement on
one of the two parties (say $A$ in the above definition)
\cite{luofu}. It is important to remark that its value is
dependent on the choice of the party to be measured \cite{notepiani}. Although in
principle this expression can be very complicated to evaluate
explicitly as it involves a minimization problem over the set of
zero-discord states, an analytical formula exists for the general
two-qubit case \cite{vedral,luofu,gprl}. In terms of the Bloch
picture [Eq.~(\ref{blopix})], one has
\[
D(\rho)=2\textrm{Tr}[S]-2\lambda_{\max}(S)\,,
\] where $\lambda_{\max}$ stands for `maximum eigenvalue', and the matrix $S$ is defined as $S=\frac14 (\vec{x} \vec{x}^T + T T^T)$.
The (square root of) geometric discord of $\rho_{\rm X}$ is then
\begin{equation}
\sqrt{D(\rho_{\rm X})}=\sqrt{[\textrm{Re}(U_{A}^{*}V_{B})]^{2}+|X|^{2}}\,.
\label{gdos}
\end{equation}
The above formula is correct up to the second order and it has an
immediate physical interpretation. The two terms in
Eq.(\ref{gdos}) come indeed from first and second order
contributions to the time evolution of the state. In particular
the $X$ term accounts for photon-exchange between the two atoms
and carries all the information available about causal propagation
and atom state dressing. Interestingly, the processes which
contribute to non-zero quantum discord are 0 and 1-photon
processes and even though the (square root of) geometric quantum
discord has a continuous evolution starting from $t=0$, it is well
sensitive to light-cone crossing, showing a peak at $t=r/c$.

\subsection{Negativity}
Negativity \cite{neg} is a well-known and easily computable
measure of entanglement for bipartite systems which is based on
the positivity of the partial transposition (PPT) criterion
\cite{ppt}. Given a general $d\otimes d$ quantum bipartite state
$\rho$, the (normalized) negativity is defined as
\begin{equation}
N(\rho)\doteq\frac{1}{d-1}||\rho^{T_{A}}-\textrm{I}_{AB}||_{1}
\label{neg}
\end{equation}
where the $T_{x}$ refers to the partial transposition operation
with respect to the $x$ party ($x=A,B$), $\textrm{I}_{AB}$ is the
identity operator in the composed Hilbert space
$\mathcal{H}_{A}\otimes\mathcal{H}_{B}$ and
$||M||_{1}=\textrm{Tr}|M|=\sum_{i}|m_{i}|$ is the trace norm for a
matrix $M$ with eigenvalues $\{m_{i}\}$. As in the case of
geometric discord we can easily compute the negativity of
$\rho_{\rm X}$ up to the second order in time-dependent
perturbation theory and we obtain the following expression
\begin{equation}
N(\rho_{\rm X})=\max\left\{0,\,\sqrt{(|U_{A}|^{2}-|V_{B}|^{2})^{2}+4|X|^{2}}-|U_{A}|^{2}-|V_{B}|^{2}\right\}\,.
\label{negus}
\end{equation}
The three physical processes contributing to entanglement are
exactly the same as for geometric discord. However, in this case
there is a time-dependent condition for entanglement to start
increasing. Indeed it is easy to check that as long as the
following condition is fulfilled
\begin{equation}
\frac{|X|^{2}}{|U_{A}|^{2}|V_{B}|^{2}}\le 1
\label{entacond}
\end{equation}
entanglement will be zero. Intuitively, this means that for
entanglement to be non-zero, second-order processes must dominate
over first-order ones. It is worth noticing that such a kind of
constraint is absent in the case of the (square root of) geometric
discord, which amounts simply to the sum of two positive and
continuous functions.

\subsection{Maximum connected correlation function}
Geometric discord and entanglement are quantities which are
strictly connected to the quantum character of a system and,
indeed, they miss a classical analogue. In this respecr, they are
key quantities when it comes to understanding the interplay
between the foundations of quantum mechanics and micro-causality,
one of the postulates of relativity theory. However, one might
also be interested in correlations arising from observable
quantities such as, for instance, angular momenta and in general
spin-like operators. We thus study the atom-atom (classical)
connected correlation functions \cite{arealaw}, to reveal the
highest level of sensitivity at which the Bloch vectors of the two
atoms perceive each other outside and inside the light cone. Given
a bipartite state $\rho$ of a pair of two-level quantum systems,
we define the maximum connected correlation function $C(\rho)$ as
follows
\begin{equation}
C(\rho)\doteq \max_{n,n'}\left\{\langle(\vec{\sigma} \cdot\hat{n})_{A}\otimes(\vec{\sigma} \cdot\hat{n}')_{B}\rangle_{\rho}-\langle(\vec{\sigma} \cdot\hat{n})_{A}\rangle_{\rho}\langle(\vec{\sigma} \cdot\hat{n}')_{B}\rangle_{\rho}\right\}\,,
\label{spin}
\end{equation}
where $\vec{\sigma}$ is the three-component Pauli-operator
vector and $(\vec{\sigma} \cdot\hat{n})$ is the projection of
such a spin vector along the direction pointed by $\hat{n}$. For
generic two-qubit states $\rho$ decomposed in Bloch form as in
Eq.~(\ref{blopix}), the maximum in Eq.~(\ref{spin}) can be
computed in closed form and reads \cite{localizable2,unpube}
\[
C(\rho)=\sqrt{\lambda_{\max}(W^T W)}\,,
\]
where $W=T-\vec{x}\, \vec{y}^T$. We have computed $C(\rho)$ for
the state $\rho_{\rm X}$ and obtained an exact expression up to
the second order,
\begin{equation}
C(\rho_{\rm X})=\max\left\{ |U_{A}|^{2}+|V_{B}|^{2}+2\textrm{Re}(A),2(|X|+|L|)\right\}
\label{spinus}
\end{equation}
where $L=U_{A}^{*}V_{B}$. Once again, in this case only 0 and
1-photon processes contribute to the above correlation function.
Moreover, with $C(\rho_{\rm X})$ being dependent on $X$, it shows
sensitivity to the light-cone crossing.

\begin{figure*}[t]
\begin{center}
\subfigure[]
{\includegraphics[height=3.8cm]{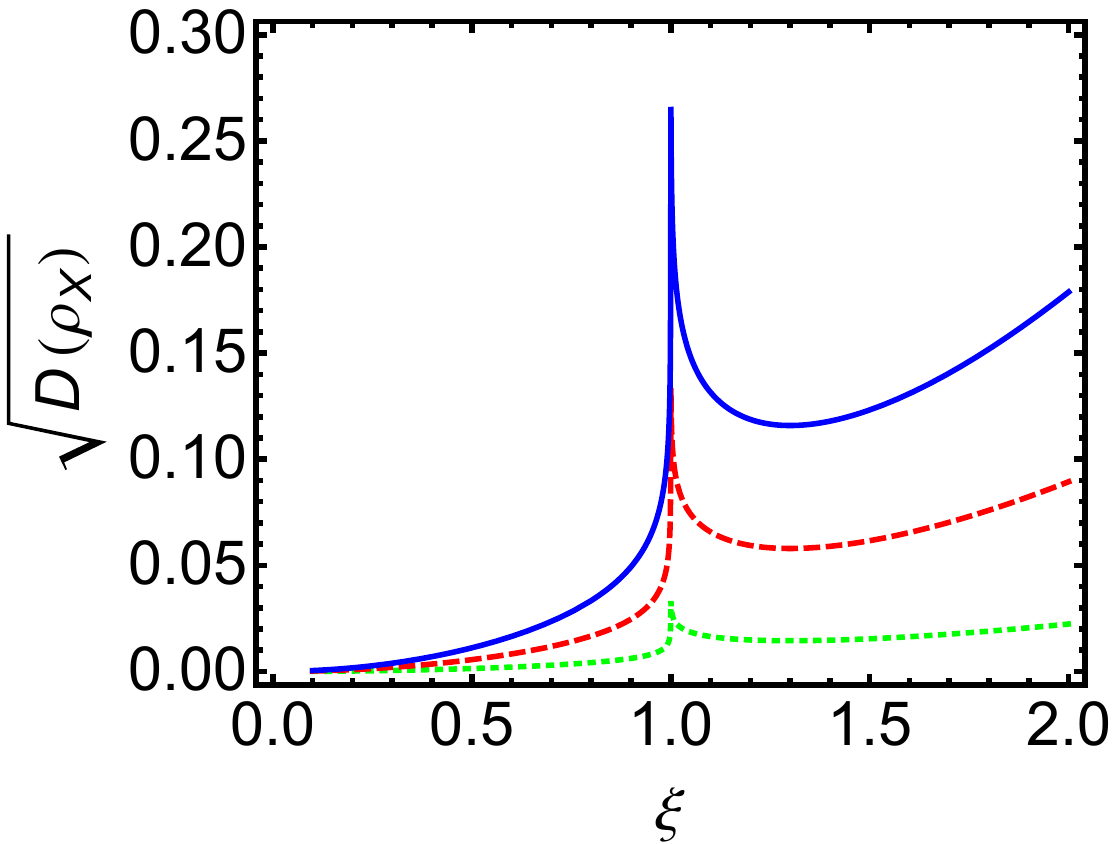}\label{geodisc}}
\hspace*{0.2cm} \subfigure[]
{\includegraphics[height=3.8cm]{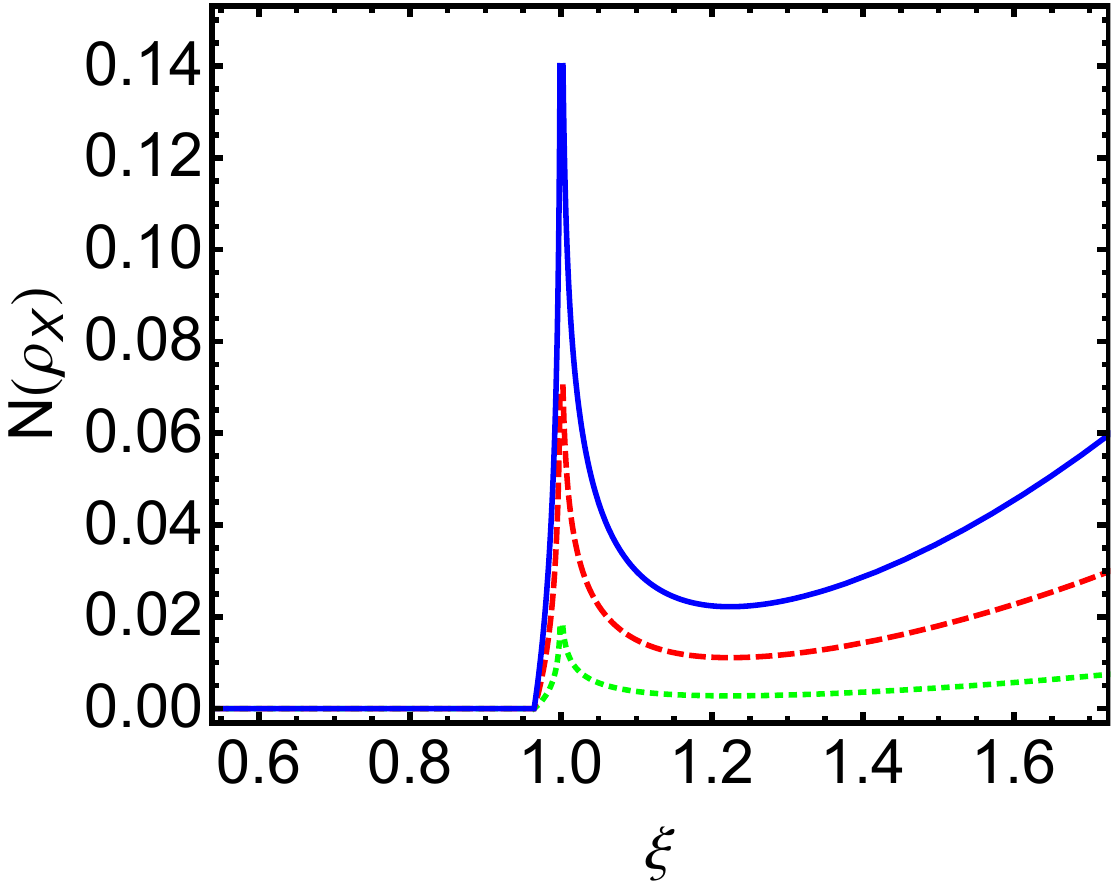}\label{negat}}
\hspace*{0.2cm} \subfigure[]
{\includegraphics[height=3.8cm]{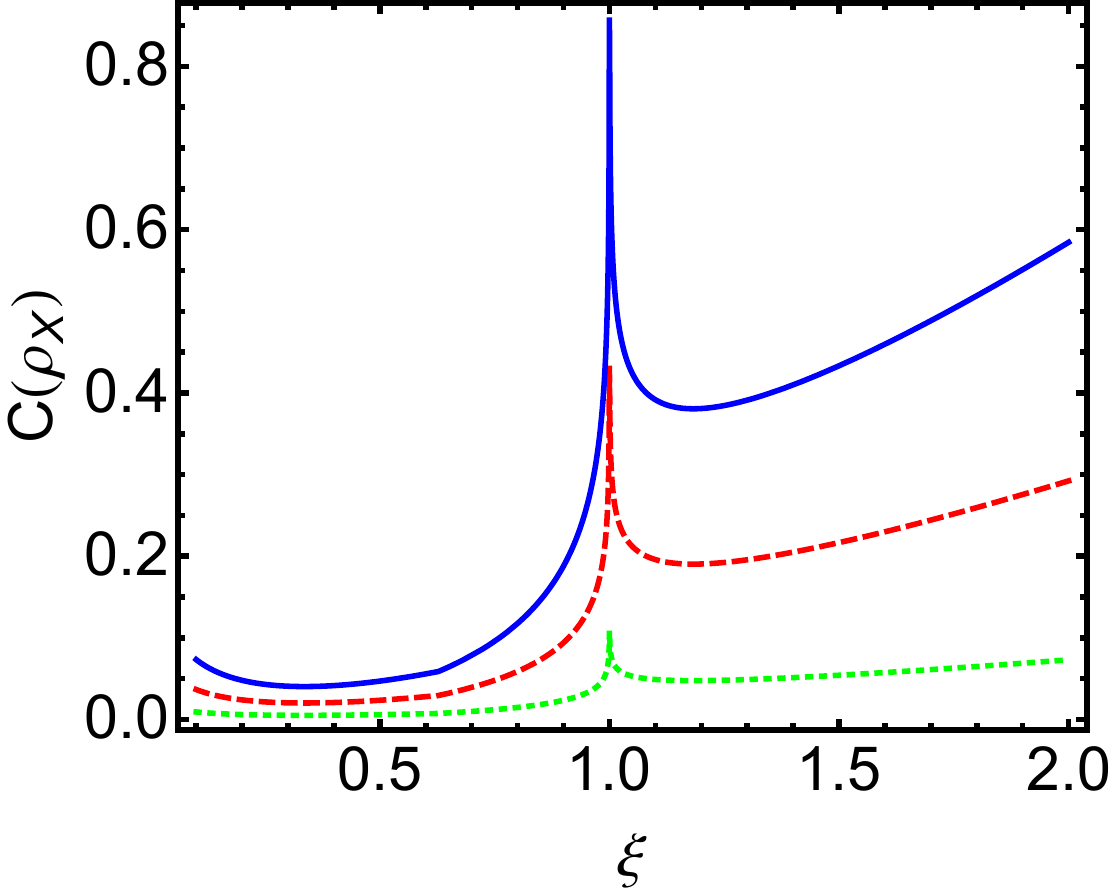}\label{corr}}
\center{\caption{(Color online) Time-evolution of (a) the square root of
geometric quantum discord $\sqrt{D(\rho_{\rm X})}$, (b) the
negativity $N(\rho_{\rm X})$ and (c) the maximum connected
correlation function $C(\rho_{\rm X})$,  for $r = \upsilon \pi / 4
\Omega$ and three different choices of the coupling strength:
$Z=50$ (dotted green), $Z=200$ (dashed red), $Z=400$ (continuous
blue).}}
\end{center}
\label{figures}
\end{figure*}


\subsection{Results and Discussion}
In the following we analyze the time evolution of the above
correlations and compare their behavior qualitatively and
quantitatively. We remark that all of the above quantities depend
on the three terms $U_{A}, V_{B}$ and $X$, while the maximum
connected correlation function displays an $A$-dependence as well.
Fig.\ref{geodisc} shows the behavior of the square root of
geometric discord $\sqrt{D(\rho_{\rm X})}$ as a function of the
re-scaled time $\xi=rt/\upsilon$ for different choices of the
atom-field coupling constant, spanning from a weak to a strong
coupling regime, and for a fixed distance $ r = \upsilon \pi / 4
\Omega$ between the qubits.

The first feature we notice, which is perhaps the most interesting
one, is the relatively slow but continuous increase that the
(square root of) geometric discord shows prior to the light-cone
crossing. A similar behavior had been previously found by one of
the present authors in \cite{carlos2} when studying the time
evolution of entanglement measured by the concurrence
\cite{concurrence} in the same model. However, unlike concurrence
which started increasing within a very short time interval just
before $\xi=1$, the (square root of) geometric discord reaches
finite values for a much longer time interval inside the
space-like region $J_S\equiv\{0\le\xi\le1\}$. If we recall the
interpretation of geometric discord mentioned above, we may argue
that a one-party measurement performed at any time inside $J_S$
will always induce a disturbance on the composite system.

Secondly, we observe a peak at $\xi=1$ which is independent of the
interaction regime. The height of such a peak, and more generally
the global magnitude of the (square root of) geometric discord,
increases as we increase the coupling. These latter features are
easily understood by looking at Eq.(\ref{gdos}). As we said above
the (square root of) geometric discord is the sum of a first order
term, which does not carry any causality-related information, and
a second order term, which instead does carry that kind of
information. Hence, the stronger the interaction is, the bigger
this second order term becomes.

In Fig.\ref{negat} we show the time evolution of the negativity
for the same three choices of coupling strengths and the same
atom-atom separation. In this case we find essentially the same
behavior as reported in \cite{carlos2} for concurrence. By
comparing Fig.\ref{geodisc} and \ref{negat}, we may conclude that
quantum discord is more sensitive to vacuum fluctuations, which
are responsible for creating correlations between the two atoms
outside the light cone. This behavior is well understood again
when one looks at Eq.(\ref{gdos}). The proportionality to $X$,
which is a second-order 0-photon term, incorporates exactly this
kind of trait.

Fig.\ref{corr} shows the time evolution of the maximum connected
correlation function $C(\rho_{\rm X})$ for the same choice of
parameters as in the previous plots. In this case we find
something very interesting and not at all easily predictable.
Indeed, like geometric discord, the maximum connected correlation
function starts increasing significantly inside $J_S$ and it shows
a peak at $\xi=1$. The maximum connected correlation function
$C(\rho)$ is not {\itshape a priori} a fully quantum quantity, and it
determines how, on average, the Bloch vectors of the two atoms influence each
other. The present results seem to suggest that this might be the
key quantity to measure when it comes to an experimental detection
of the dynamics of correlations, provided that a simultaneous set
of optimal measurements on the two qubits can be efficiently
performed in the laboratory frame.

\begin{figure}[tb]
\begin{center}
\includegraphics[width=10cm]{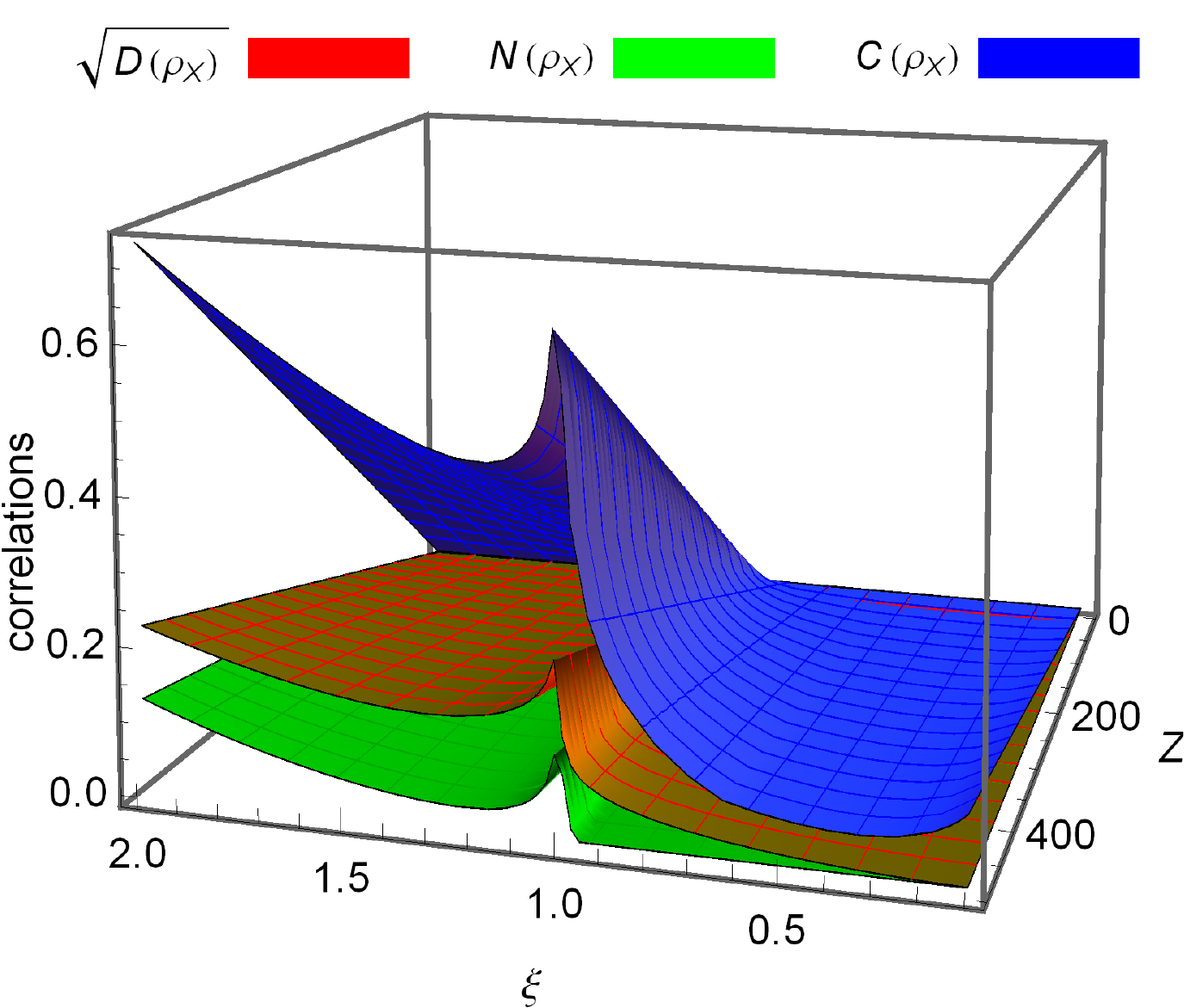}
\caption{(Color online) Comparative plot displaying the maximum
connected correlation function $C$ (topmost surface, blue online),
the square root $\sqrt{D}$ of the geometric discord (middle
surface, red online), and the negativity $N$ (bottommost surface,
green online), calculated for the state $\rho_X$ as functions of
the dimensionless time $\xi$ and of the coupling strength $Z$, for
$r = \upsilon \pi / 4 \Omega$.} \label{all}
\end{center}
\end{figure}

It is worth noticing here that the optimal ``measurement
directions'' $\hat n, \hat n'$ are completely different in the
space-like region $J_S$ and on the light cone. Indeed, for $\xi <
1$, the correlation between the two Bloch vectors is best
highlighted by measuring the effective spin projections in the
equatorial $x-y$ plane. On the light cone, on the other hand, the
best choice is to measure $\sigma_z$ for both qubits. This appears
to be related to the fact that no excitation can reach the atom
$B$ before $\xi =1$ and that, as demonstrated in \cite{carlos2},
it is only after this time that the excited state population of
atom $B$ starts depending on the presence of atom $A$. The vacuum
fluctuations, thus, are able to correlate essentially transversal
observables for $\xi <1$, while for a longitudinal ($z-z$)
correlation, one has to wait the arrival of the light signal.
Finally, in Fig. \ref{all} reports a visual comparison of all the
three indicators of correlations considered in the present analysis,
as functions of $Z$ and $\xi$. As anticipated, there is no
violation of the general hierarchy (\ref{hierardy}), thus
confirming that the perturbative analysis we have performed is
consistent up to the present expansion order.

\section{Non-locality}\label{inequality}
\begin{figure*}[t]
\begin{center}
\subfigure[]
{\includegraphics[width=5.7cm]{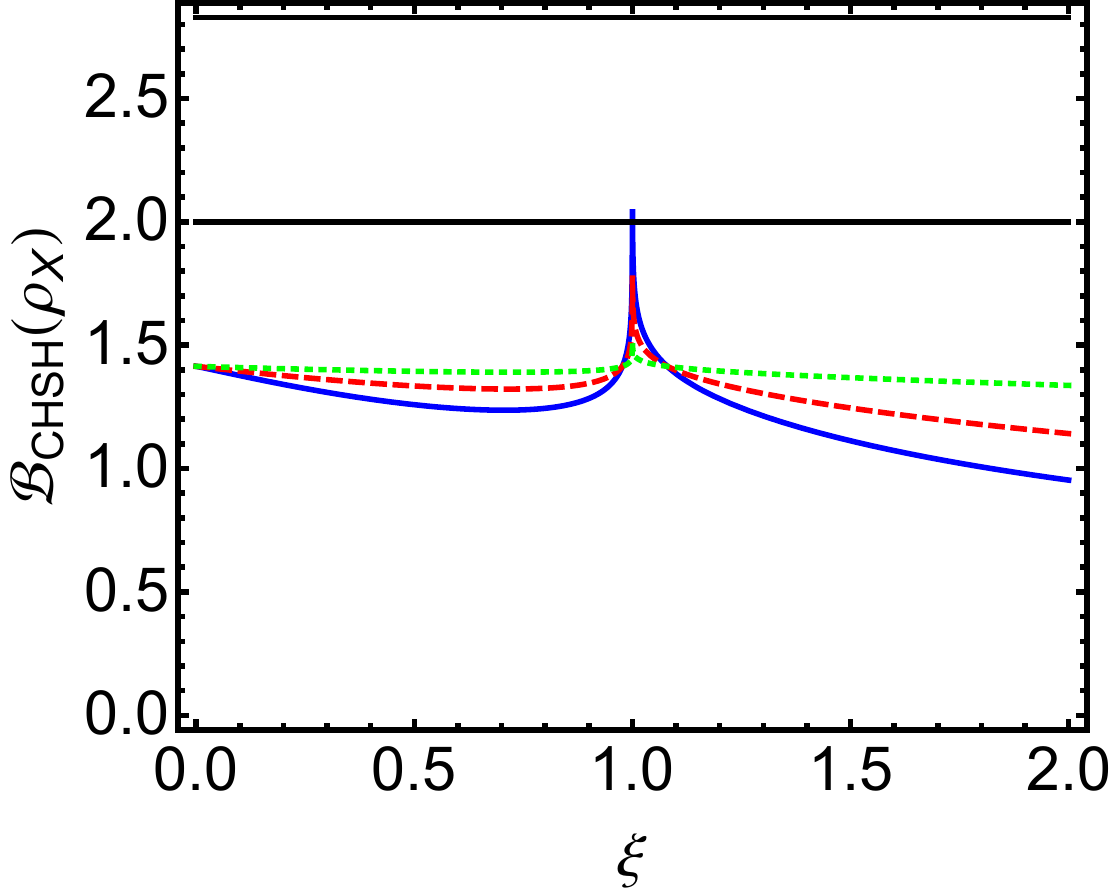}\label{bell}}
\hspace*{0.2cm} \subfigure[]
{\includegraphics[width=5.7cm]{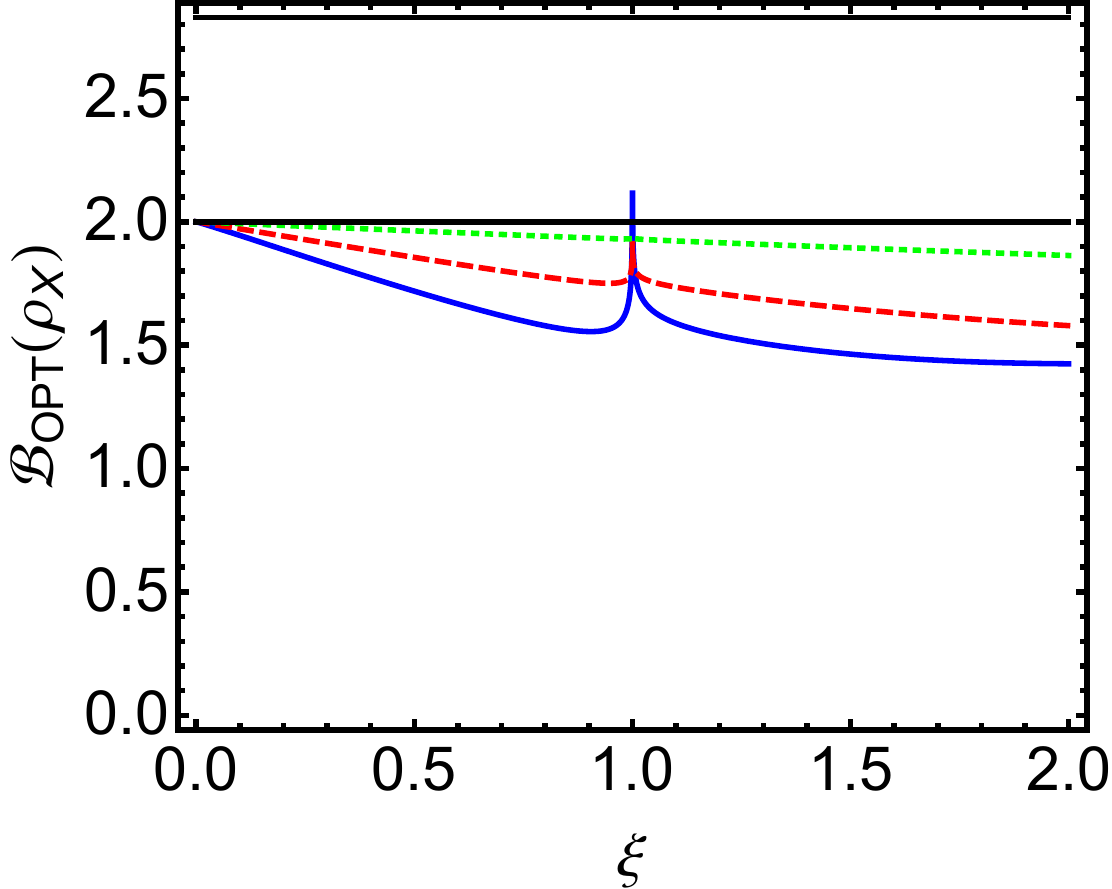}\label{opt}}
\hspace*{0.2cm} \subfigure[]
{\includegraphics[width=5.7cm]{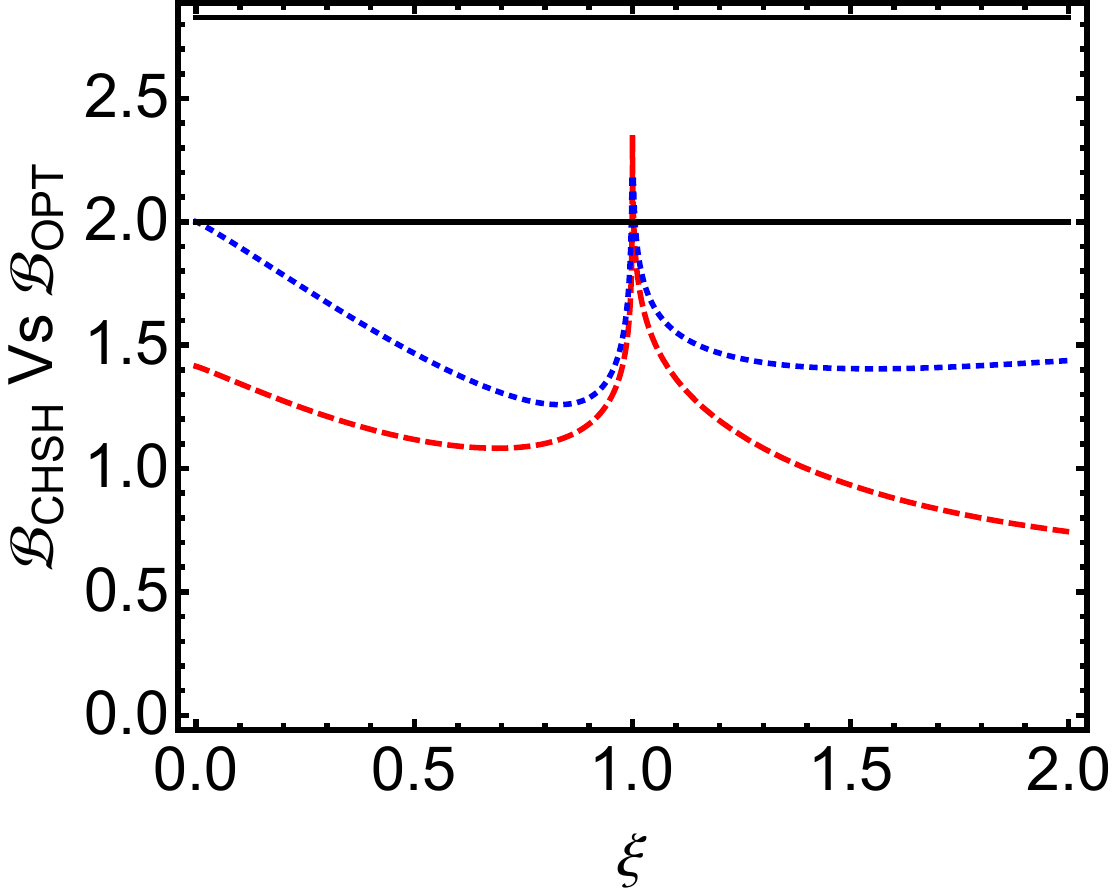}\label{confr}}
\caption{(Color online) (a) Bell-CHSH parameter ${\cal B}_{CHSH}$ and (b) its optimized version ${\cal B}_{\textrm{OPT}}$, plotted for  $Z =
100$ (dotted green), $Z = 400$ (dashed red) and $Z =
800$ (continuous blue), as a function of $\xi$. The straight line at ${\cal B}=2$ gives the limit for
a local realistic description.
 As one can see, in both cases, a very strong
coupling is needed in order to have a clear violation of the
inequality, occurring, however, only very close to the light cone
crossing point.
(c) Comparison between
${\cal B}_{CHSH}$ (red dashed) and ${\cal B}_{\textrm{OPT}}$ (blue dotted) for strong coupling $Z=1000$. In all these plots $r = \upsilon \pi / 4
\Omega$.
} \label{chsh}
\end{center}
\label{figures}
\end{figure*}

One may wonder what the above results mean in term of
non-locality. In this respect, a key quantity to investigate
non-local effects in the dynamics of two two-level systems is the
Bell parameter \cite{bell}, resulting from well-known inequalities
that local classical hidden variable theories cannot violate. To
make a long story short, one identifies a set of joint
measurements to be performed on the composite system. Then, based
on the outcomes of such measurements, it is possible to define a
statistical parameter $\cal{B}$ for which a classical threshold
value $\cal{B}_{C}$ exists. Whenever the inequality
$$
\cal{B}<\cal{B}_{C}
$$
is violated, the state of the system under scrutiny is not
reproducible by means of local classical hidden variable theories.
Unfortunately, the other way around is not true: some mixed
entangled states exist that do not violate any Bell inequality
\cite{werner}. In the case of two two-level system, it is well
known that $\cal{B}_{C}=$ 2, whereas the maximum violation allowed
by quantum mechanics is given by the so-called Tsirelson bound,
$\cal{B}_{\max}=$ 2$\sqrt{2}$, which is saturated by maximally
entangled  states. Hence, if  for a given bipartite quantum state
$\rho$ we find $2<\cal{B}(\rho)\leq 2\sqrt{2}$, such a state is
not classically reproducible. We have considered two different
Bell parameters in the present analysis: the conventional CHSH one
\cite{chsh} and its optimized version for ``X''-shaped states such
as $\rho_{\rm X}$, given in \cite{bellomo}. The former, for the
states in Eq.~(\ref{state}), reads as follows (up to the second
perturbative order)
\begin{equation}
{\cal B}_{CHSH}(\rho_{\rm
X})=-\sqrt{2}(\rho_{11}+\rho_{44}-\rho_{22}-\rho_{33}+2\textrm{Re}\rho_{23}+2\textrm{Re}\rho_{14})
\, , \label{chsh}
\end{equation}
whereas the latter is \cite{bellomo}
\begin{equation}
{\cal B}_{OPT}(\rho_{\rm
X})=2\sqrt{u_{1}+\max[u_{2},u_{3}]}\, ,  \label{bello}
\end{equation}
where
$$
u_{1}=4(|\rho_{14}|+|\rho_{23}|)^{2}\qquad u_{3}=4(|\rho_{14}|-|\rho_{23}|)^{2}
$$
$$
u_{2}=(\rho_{11}+\rho_{44}-\rho_{22}-\rho_{33})^{2} \, .
$$
The above quantities correspond to two different choices of the
Bell parameters, that is, different choices of the angles along
which we project the effective spin operators in a
joint-measurement experiment. The main difference between them is
that ${\cal B}_{OPT}$ is optimal in the sense that it
maximizes the violation of the related Bell inequality, whenever
such a violation is present. We report in Fig.\ref{bell} the time
evolution of the the Bell parameter ${\cal B}_{CHSH}$. It is clear
from these plots that, in order to observe a violation of the Bell
inequality, a very strong coupling is required ($Z\approx 1000$).
However, such a violation is witnessed only in the surroundings of
the light cone crossing $\xi=1$. Fig.\ref{opt} shows, instead, the
time evolution of ${\cal B}_{OPT}$. Some qualitative and
quantitative differences between these quantities, especially for
$\xi>1$ and very strong coupling, are present. First of all we
notice that, as expected, the optimized Bell parameter ${\cal B}_{OPT}$ is greater than ${\cal B}_{CHSH}$ for all
couplings and times. Secondly, the latter is clearly more
sensitive to the strong-coupling regime. This result makes perfectly
sense if we look at the entanglement dynamics as a function of the
coupling constant: the bigger $Z$ is, the more entanglement is
present in the system, pushing up the Bell parameters' value.
These last two features are best enlightened in the third plot
\ref{confr} where the dynamics of the two Bell parameters we
considered, is compared in the case of very strong coupling
$Z=1000$.

\section{Experimental implementation}
\label{expeimple} Now we will focus on the particular experimental
set-up that we propose to test the results above. Clearly, we need
the ability of fast on-off switching of the qubit-field
interaction. We should be able to prepare the initial state Eq.
(\ref{eq:initialstate}) and let the interaction be active only
during a finite time. We can conceive several circuit QED schemes
in order to achieve these goals. As commented before our general
formalism can accommodate several architectures; in particular, it
is valid both for inductive and capacitive couplings. We can
choose for instance a pair of three-junction flux qubits
galvanically coupled to the center conductor of an open
transmission line. In Ref. \cite{carlos1} it is shown how the
desired initial state can be prepared with high fidelity by
varying an external magnetic flux adiabatically for qubit $B$ and
non-adiabatically for qubit $A$. After that, the interaction has
to be switched on and kept constant during a given time interval.
In Ref. \cite{peropadre} several modifications of the
three-junction scheme are proposed in order to achieve couplings
tunable in strength up to the ultra-strong regime. In particular,
a specific set-up featuring an intermediate superconducting loop
has been described in details in Ref. \cite{carlos3}, where an
effective interaction Hamiltonian between a qubit and a
transmission line has been derived, that reads
\begin{equation}
H_I \propto \cos(f)\sigma_x\,V(x), \label{effhamilt}
\end{equation}
$f$ being related with an external magnetic flux $\Phi$ threading
the SQUID: $f=2\pi\Phi/\Phi_0$ (where $\Phi_0=h/2e$ is the flux
unit). By a suitable modulation of the flux $\Phi$, thus, the
interaction can be activated and switched off at will. State of
the art circuit-QED technology allows variations of the magnetic
flux at frequencies of about $10\,\mbox{GHz}$ \cite{casimirwilson}
and larger values are expected for the future \cite{guille}. As a
result, switching times of the order of $0.1\, ns$ and even
shorter can be safely considered. Taking the values that we have
considered in our plots, and typical circuit-QED parameters, such
as $\upsilon=1.2\cdot10^8\, \mbox{m/s}$ and $\Omega=10^9 \,
\mbox{m/s}$, the point $\xi=1$ is equivalent to an interaction
time $t \simeq 1\, \mbox{ns}$. Thus, the region around $\xi=1$ is
well within experimental reach. In particular the strongest value
that we considered for the adimensional coupling $Z$  is
equivalent to $g/\Omega \simeq 0.3$, which is quite similar to the
ones in cutting-edge experiments investigating the ultra-strong
coupling regime \cite{forn-diaz10,niemczyk10}.

Once the interaction is switched off, quantum state tomography
\cite{tomographycircuits} may be performed in order to quantify
the degree of correlations, using for instance the magnitudes that
we have considered in this work. The dynamics is effectively
frozen and the system remains in the state $\rho_{\rm X} (t)$, so
measurements can take as much time as required. In particular, it
would be interesting to run the experiment for different
interaction times inside and outside the light cone, in order to
test both the peak at $\xi=1$ and the correlations for $\xi<1$.

\section{Concluding Remarks}
\label{conclu} To summarize, using second order perturbation
theory we have discussed the dynamics of quantum correlations in
the Fermi problem, which can be experimentally tested in a
one-dimensional set-up involving two artificial atoms coupled to
the electromagnetic field of an open ended transmission line. We
have compared the time behavior of the entanglement, as measured
by the negativity, and of more general quantum correlations such
as the (square root of) geometric discord, and the maximum
connected correlation function. All of these correlations display
a peak at the light cone crossing point, $\xi =1$, which
corresponds to the time at which a signal from atom $A$ arrives at
$B$. The geometric discord and the connected correlation, however,
have a substantially non-zero value also in the space-like region.
This is due to the fact that electromagnetic vacuum fluctuations
can induce (transverse) correlations that are signalled by these
functions. As the light-cone is crossed, these correlations change
their character and become longitudinal as a result of the fact
that the excited state population of the second atom starts to
depend on the presence of the first one. We have briefly
investigated non-local effects in this model as encoded in
possible violations of Bell inequalities. We have found that a
violation can occur in the neighborhood of the light-cone crossing
and only for strong couplings between the atoms and the
propagating field. We
believe that both geometric discord and maximum connected
correlations, which are found to be sensitive to causal
propagation, are suitable candidates for understanding and testing
experimentally the role of micro-causality in the dynamics of
quantum correlations.

\section*{Acknowledgments}{We acknowledge the University of Nottingham
(ECRKTA/2011), the U.K. EPSRC [Grants EP/J016349/1 and EP/J016314/1 (subcode RDF/BtG/0612b/31)], the Finnish Cultural Foundation [Science Workshop on Entanglement], and the Emil Aaltonen foundation [Non-Markovian Quantum Information] for financial support. GA thanks D Girolami and M Piani for discussions.}


\section*{References}
\begin{thebibliography}{80}
\bibitem{fermi} Fermi E 1932 {\it Rev. Mod. Phys.} {\bf 4} 87

\bibitem{biswas} Biswas A K, Compagno G, Palma G M, Passante R and Persico F 1990 {\it Phys. Rev. A} {\bf 42} 4291

\bibitem{heger} Hegerfeldt G C 1994 {\it Phys. Rev. Lett.} {\bf 72} 596

\bibitem{buch} Buchholz D and Yngvason J 1994  {\it Phys. Rev. Lett.} {\bf 73} 613

\bibitem{power} Power E A and Thirunamachandran T 1997 {\it Phys. Rev. A} {\bf 56} 3395

\bibitem{milonni} Milonni P W, James D F V and Fearn H 1995 {\it Phys. Rev. A} {\bf 52} 1525

\bibitem{carlos1} Sab{\'i}n C, del Rey M, Garc\'ia-Ripoll J J and Le\'on J 2011 {\it Phys. Rev. Lett.} {\bf 107} 150402

\bibitem{carlos2} Sab{\'i}n C, Garc\'ia-Ripoll J J, Solano E and Le\'on J 2010 {\it Phys. Rev. B} {\bf 81} 184501

\bibitem{concurrence} Hill S and Wootters W K 1997 {\it Phys. Rev. Lett.} {\bf 78} 5022

\bibitem{zurek}
 Ollivier H and Zurek W H 2001 {\it Phys. Rev. Lett.} {\bf 88} 017901

\bibitem{anvedi}
 Henderson L and Vedral V 2001 {\it J. Phys. A} {\bf 34} 6899

 \bibitem{ferraro}
Ferraro A, Aolita L, Cavalcanti D, Cucchietti F M and Ac\'in A
2010 {\it Phys. Rev. A} {\bf 81} 052318

\bibitem{piani} Piani M, Gharibian S, Adesso G, Calsamiglia J, Horodecki P and Winter A 2011 {\it Phys. Rev. Lett.} {\bf 106} 220403
\bibitem{streltsov}
 Streltsov A, Kampermann H and Bruss D 2011 {\it Phys. Rev. Lett.} {\bf 106} 160401

\bibitem{plastica}
Campbell S, Apollaro T J G, Di Franco C, Banchi L, Cuccoli A, Vaia R, Plastina F and Paternostro M 2011 {\it  Phys. Rev. A} {\bf 84} 052316

\bibitem{gerardo} Girolami D and Adesso G 2011 {\it Phys. Rev. A} {\bf 84} 052110

\bibitem{pianinew}
Piani M and  Adesso G 2012 {\it Phys. Rev. A} {\bf 85} 040301(R)

\bibitem{reviewmodi}
Modi K, Brodutch A, Cable H, Paterek T and Vedral V 2011 {\it e--print}  arXiv:1112.6238

\bibitem{luo}  Luo S 2008 {\it Phys. Rev. A} {\bf 77} 042303

\bibitem{davide} Girolami D and  Adesso G 2011 {\it Phys. Rev. A} {\bf 83} 052108

\bibitem{discordgaussiano} Adesso G and Datta A 2010 {\it Phys. Rev. Lett.} {\bf 105} 030501;
Giorda P and Paris M G A 2010 {\it Phys. Rev. Lett.} {\bf 105} 020503

\bibitem{vedral}
Daki\'c B, Brukner C and Vedral V 2010 {\it  Phys. Rev. Lett.} {\bf 105} 190502

\bibitem{kavan}
Modi K, Paterek T, Son W, Vedral V and Williamson M 2010 {\it Phys. Rev. Lett.} {\bf 104} 080501

\bibitem{localizable1}
Verstraete F, Popp M and Cirac J I 2004 {\it Phys. Rev. Lett.} \textbf{92} 027901

\bibitem{localizable2}
Popp M, Verstraete F, Mart\'in-Delgado M A and Cirac J I 2005 {\it Phys. Rev. A} {\bf 71} 042306

\bibitem{arealaw}
Wolf M M, Verstraete F, Hastings M B and Cirac J I 2008 {\it Phys. Rev. Lett.} {\bf 100} 070502

\bibitem{neg} Vidal G and Werner R F 2002  {\it Phys. Rev. A} {\bf 65} 032314

\bibitem{unpube} Adesso G (unpublished)

\bibitem{luofu} Luo S and Fu S 2010 {\it Phys. Rev. A} {\bf 82} 034302

\bibitem{gprl} Girolami D and Adesso G 2012 {\it Phys. Rev. Lett.} {\bf 108} 150403

\bibitem{notepiani} The geometric discord $D$ can increase under quantum operations on the party that is not measured \cite{cinesidigo, pianicomment}; as such, it should be regarded just as an indicator of nonclassical correlations rather than as a proper measure. It constitutes nonetheless a valid lower bound to another {\it bona fide} distance-based measure of quantum correlations defined in terms of relative entropy \cite{piani,kavan}, and it enjoys a specific operational interpretation for two-qubit states \cite{dakicrsp}.


\bibitem{cinesidigo}
Hu X, Fan H, Zhou D L and Liu W-M 2012 {\it e--print} arXiv:1203.6149

\bibitem{pianicomment}
Piani M 2012 {\it e--print} arXiv:1206.0231


\bibitem{dakicrsp}
Daki\'c B {\it et al} 2012 {\it e--print}  arXiv:1203.1629


\bibitem{ppt} Peres A 1996 {\it Phys. Rev. Lett.} {\bf 77} 1413; Horodecki R, Horodecki P and Horodecki M 1996 {\it Phys. Lett. A} {\bf 210} 377

\bibitem{bell} Bell J S 1964 {\it Physics} {\bf 1} 195


\bibitem{werner}
Werner R F 1989 {\it Phys. Rev. A} {\bf 40} 4277


\bibitem{chsh} Clauser J F, Horne M A, Shimony A and Holt R A 1969 {\it Phys. Rev. Lett.} {\bf 23} 880

\bibitem{bellomo} Bellomo B, Lo Franco R and Compagno G 2010 {\it  Phys. Lett. A} {\bf 374} 3007

\bibitem{casimirwilson} Wilson C M, Johansson G, Pourkabirian A, Johansson J R,  Duty T, Nori F, and Delsing P 2011 {\it Nature} {\bf 479} 376

\bibitem{peropadre} Peropadre B, Forn-D{\'i}az P, Solano E and Garc{\'i}a-Ripoll J J 2010 {\it Phys. Rev. Lett.} {\bf 105} 023601

\bibitem{carlos3} Sab{\'i}n C, Peropadre B, del Rey M and Mart{\'i}n-Mart{\'i}nez E 2012 {\it e--print} arXiv:1202.1230

\bibitem{guille} Romero G, Ballester D, Wang Y M, Scarani V and Solano E 2012 {\it Phys. Rev. Lett.} {\bf 108}  120501

\bibitem{forn-diaz10} Forn-D{\'i}az P,  Lisenfeld J, Marcos D, Garc{\'i}a-Ripoll J J, Solano E, Harmans C J P M and Mooij J E 2010 {\it Phys. Rev. Lett.} {\bf 105} 237001

\bibitem{niemczyk10} Niemczyk T {\it et al} 2010 {\it Nature Phys.} {\bf 6} 772

\bibitem{tomographycircuits} Steffen M {\it et al} 2006 Science {\bf 313} 1423

\end{thebibliography}
\end{document}